\documentclass[12pt, a4paper]{article} % A4 size
\usepackage{hyperref}
\hypersetup{pdfstartview = FitH,
            pdfborder    = {0 0 0.5},
            colorlinks   = false
}
 \usepackage{epsfig, subcaption}
 \newcommand{\rmXA}      [2] {\sp{#2} {\rm #1}}
\newcommand{\mchi}           {m_{\chi}}
\newcommand{\sigmaSI}        {\sigma_0^{\rm SI}}
\newcommand{\sigmaSD}        {\sigma_0^{\rm SD}}
\newcommand{\sigmapSI}       {\sigma_{\chi {\rm p}}^{\rm SI}}
\newcommand{\armp}           {a_{\rm p}}
\newcommand{\armn}           {a_{\rm n}}
\newcommand{\FSIQ}           {F_{\rm SI}^2(Q)}
\newcommand{\FSDQ}           {F_{\rm SD}^2(Q)}
\newcommand{\Qmax}           {Q_{\rm max}}
\newcommand{\vchiLab}        {v _{\chi, {\rm Lab}}}
\newcommand{\thetaNRchi}     {\theta_{\rm N_R, \chi_{in}}}
 \newcommand{\rmF}           {\rmXA{F}  {19}}
 \newcommand{\rmAr}          {\rmXA{Ar} {40}}
 \newcommand{\rmGe}          {\rmXA{Ge} {73}}
 \newcommand{\rmXe}          {\rmXA{Xe}{129}}
 \newcommand{\rmW}           {\rmXA{W} {183}}
\newcommand{\InsertPlotNQ} [2] [t!] {
\begin{figure} [#1]
\begin{center}
 \begin{subfigure} [c] {8.25 cm}
  \includegraphics [width = 8.25 cm] {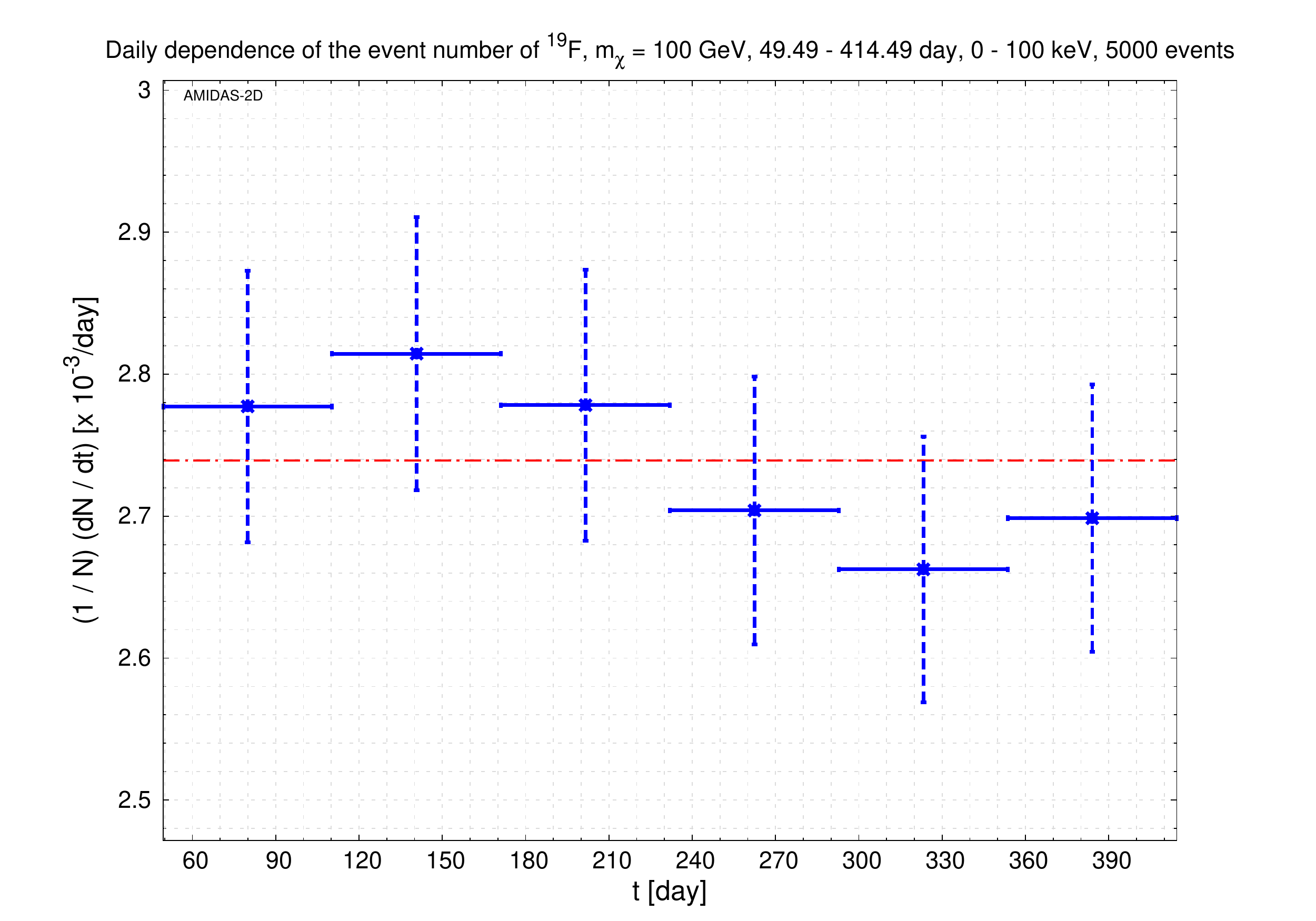}
 \caption{Event number}
 \end{subfigure}
 \hspace{0.1 cm}
 \begin{subfigure} [c] {8.25 cm}
  \includegraphics [width = 8.25 cm] {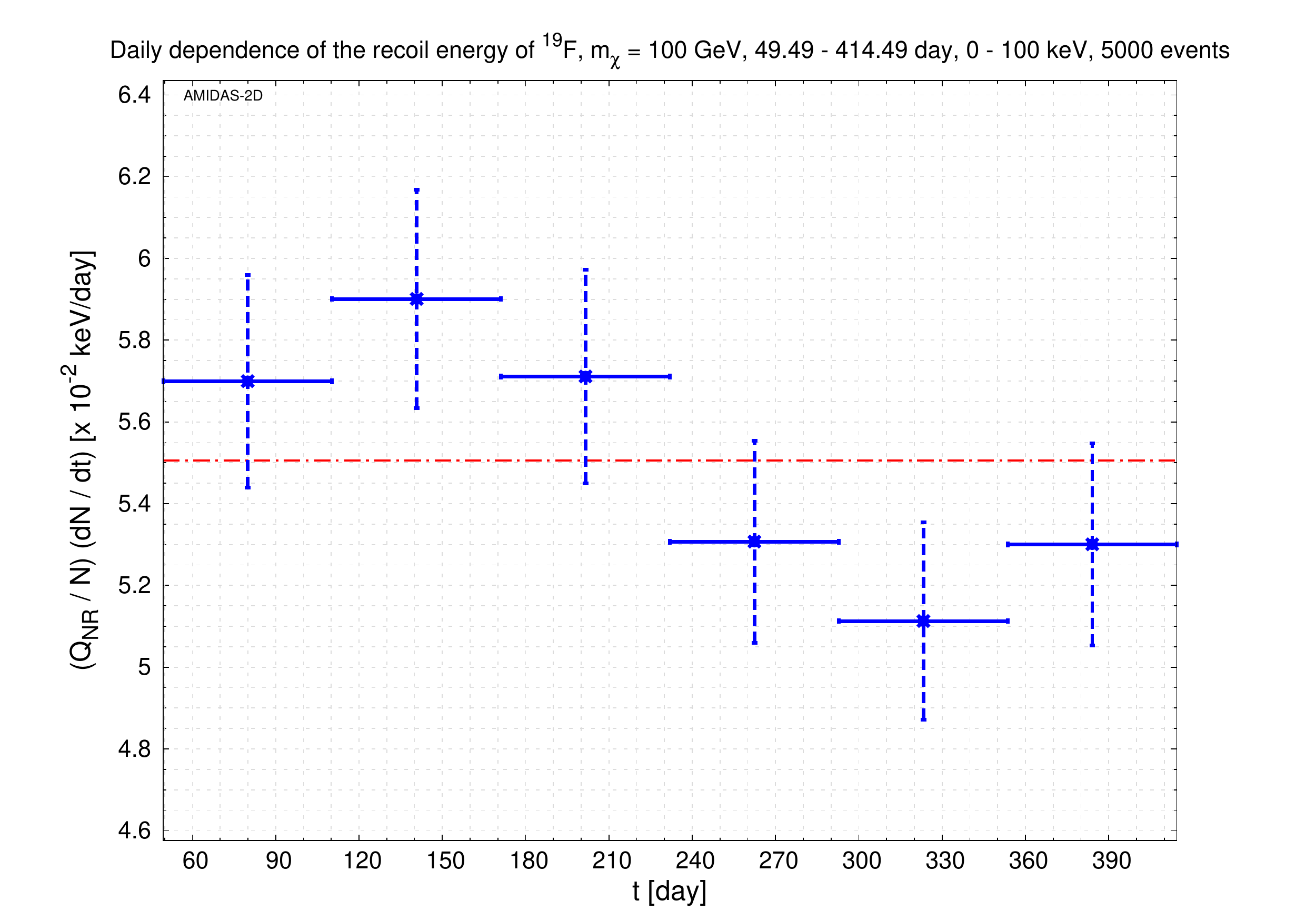}
 \caption{Accumulated recoil energy}
 \end{subfigure}
\end{center}
\caption{
 #2
}
\label{fig:NQ-\Target-\WIMPmass-5000-04949-00}
\end{figure}
}
\begin{document}
\thispagestyle{empty}
\begin{flushright}
 April 2022
\end{flushright}
\begin{center}
{\LARGE\bf
 Normal and Reverse Annual Modulations of              \\ \vspace{0.4 cm}
 Elastic WIMP--Nucleus Scattering Signals}             \\
\vspace*{0.7 cm}
 {\sc Chung-Lin Shan}                                  \\
\vspace{0.5 cm}
 {\small\it
  Preparatory Office of
  the Supporting Center for
  Taiwan Independent Researchers                       \\ \vspace{0.05 cm}
  P.O.BOX 21 National Yang Ming Chiao Tung University,
  Hsinchu City 30099, Taiwan, R.O.C.}                  \\~\\~\\
 {\it E-mail:} {\tt clshan@tir.tw}
\end{center}
\vspace{2 cm}
\begin{abstract}

 Following our earlier work on
 the 3-dimensional effective velocity distribution of
 Galactic WIMPs
 (not only impinging on our detectors
  but also)
 scattering off target nuclei,
 in this paper,
 we demonstrate
 the normal and a ``reverse'' annual modulations of
 elastic WIMP--nucleus scattering signals,
 which could be observed
 in direct Dark Matter detection experiments.
 Our simulations show that,
 once the WIMP mass is as light as only a few tens GeV,
 the event number and the accumulated recoil energy of
 WIMP--induced scattering events off
 both of light and heavy target nuclei
 would indeed be maximal (minimal)
 in summer (winter).
 However,
 once the WIMP mass is as heavy as a few hundreds GeV,
 the event number and the accumulated recoil energy of
 WIMP scattering events off heavy nuclei
 would inversely be minimal
 in summer.
 Understandably,
 for an intermediate WIMP mass,
 the event number and the accumulated recoil energy of
 scattering events off
 some middle--mass nuclei
 would show an approximately uniform time dependence.

\end{abstract}
\clearpage
\section{Introduction}

 Direct Dark Matter (DM) detection experiments
 aiming to observe scattering signals of
 Weakly Interacting Massive Particles (WIMPs)
 off target nuclei
 by measuring nuclear recoil energies
 deposited in an underground detector
 would still be the most reliable experimental strategy
 for identifying Galactic DM particles
 and determining their properties
 \cite{SUSYDM96, Schumann19, Baudis20, Cooley21}.
 Considering the orbital motion of the Earth around the Sun,
 the relative velocity of the Earth/laboratory
 to Galactic halo
 varies annually,
 and so the 3-dimensional velocity distribution of incident halo WIMPs
 impinging on our detectors
 \cite{DMDDD-N, DMDDD-fv_eff}.
 Hence,
 the annual modulation of the event rate for
 elastic WIMP--nucleus scattering
 has been proposed for more than three decades
 as a useful experimental strategy
 for discriminating annually varying WIMP signals
 from theoretically uniform backgrounds
 \cite{Freese88}.
 In practice,
 DAMA Collaboration claims their positive observations
 in the last two decades
 \cite{Bernabei03c, Bernabei14a, Bernabei21},
 which can however not be confirmed by other collaborations yet
 \cite{Froborg16, deSouza16, GAdhikari19a, Amare21}.

 Meanwhile,
 in standard material about
 the annual modulation of WIMP scattering event rate,
 only the variation of the WIMP incident flux
 proportional to its incoming velocity
 in the laboratory reference frame
 has been taken into account
 \cite{Freese88,
       SUSYDM96, Schumann19, Baudis20, Cooley21}.
 Nevertheless,
 in our study on
 the 3-dimensional effective velocity distribution of
 Galactic WIMPs
 (not only impinging on our detectors
  but also)
 scattering off target nuclei
 \cite{DMDDD-fv_eff},
 it was demonstrated that,
 in addition to the incident flux,
 the scattering cross section (nuclear form factor) suppression
 could also affect the scattering probability of incident halo WIMPs
 moving with different velocities,
 especially
 in high recoil energy range
 as well as
 when the target nucleus and WIMPs are heavy.

 More precisely,
 while
 a WIMP moving with a higher incoming velocity
 can pass more target nuclei
 (in a unit time)
 and thus have more opportunities
 to scatter off one of them,
 its larger kinetic energy
 which could transfer to the scattered nucleus
 would in contrast reduce the scattering probability
 due to the nuclear form factor suppression
 \cite{DMDDD-fv_eff},
 especially
 when heavy target nuclei are used
 and WIMPs are also heavy.
 Additionally,
 we demonstrated in Ref.~\cite{DMDDD-fv_eff} that
 the 3-D WIMP effective velocity distribution
 show two opposite angular distribution patterns
 for light and heavy target nuclei,
 respectively,
 once the mass of incident WIMPs
 is as heavy as a few hundreds GeV.

 These interesting discoveries inspired us
 to ask whether
 the elastic WIMP--nucleus scattering event rate
 would also vary annually
 in two opposite ways
 for light and heavy target nuclei,
 in particular,
 for the case of heavy WIMPs.
 Therefore,
 in this paper,
 we would like to
 apply our simulation package for
 3-D elastic WIMP--nucleus scattering
 described in Ref.~\cite{DMDDD-3D-WIMP-N}
 to study the annual variation of
 the event number of recorded WIMP scattering signals.

 The remainder of this paper is organized as follows.
 In Sec.~2,
 we describe briefly our theoretical considerations
 regarding the annual variation of
 WIMP--nucleus scattering signals.
 Then
 three different types of
 the annual modulation
 depending on the WIMP mass and the target nucleus
 will be presented in Sec.~3.
 We summarize our observations
 in Sec.~4.

%
% 1/4
 %
%
\section{What do we know so far -- theoretical considerations}
\label{sec:considerations}

 In this section,
 we summarize our theoretical considerations
 regarding the annual variation of
 WIMP--nucleus scattering,
 which we know so far.

\subsubsection*{Flux proportionality to the WIMP incident velocity}
\label{sec:flux_v}

 As mentioned in Introduction,
 in standard material for direct Dark Matter detection physics,
 the annual modulation of the WIMP scattering event rate
 is caused by
 the yearly variation of the WIMP flux
 proportional to the WIMP incident velocity
 due to the Earth's orbital motion around the Sun
 \cite{Freese88,
       SUSYDM96, Schumann19, Baudis20, Cooley21}.
 This is easy to understand:
 the higher (lower) the incident velocity of halo WIMPs,
 the more (fewer) the target nuclei
 passed by an incident WIMP
 (in a unit time)
 and thus
 the larger (smaller)
 the scattering opportunity.

 However,
 ...

\subsubsection*{Cross section (nuclear form factor) suppression}
\label{sec:FQ}
\begin{figure} [t!]
\begin{center}
 \begin{subfigure} [c] {8.25 cm}
  \includegraphics [width = 8.25 cm] {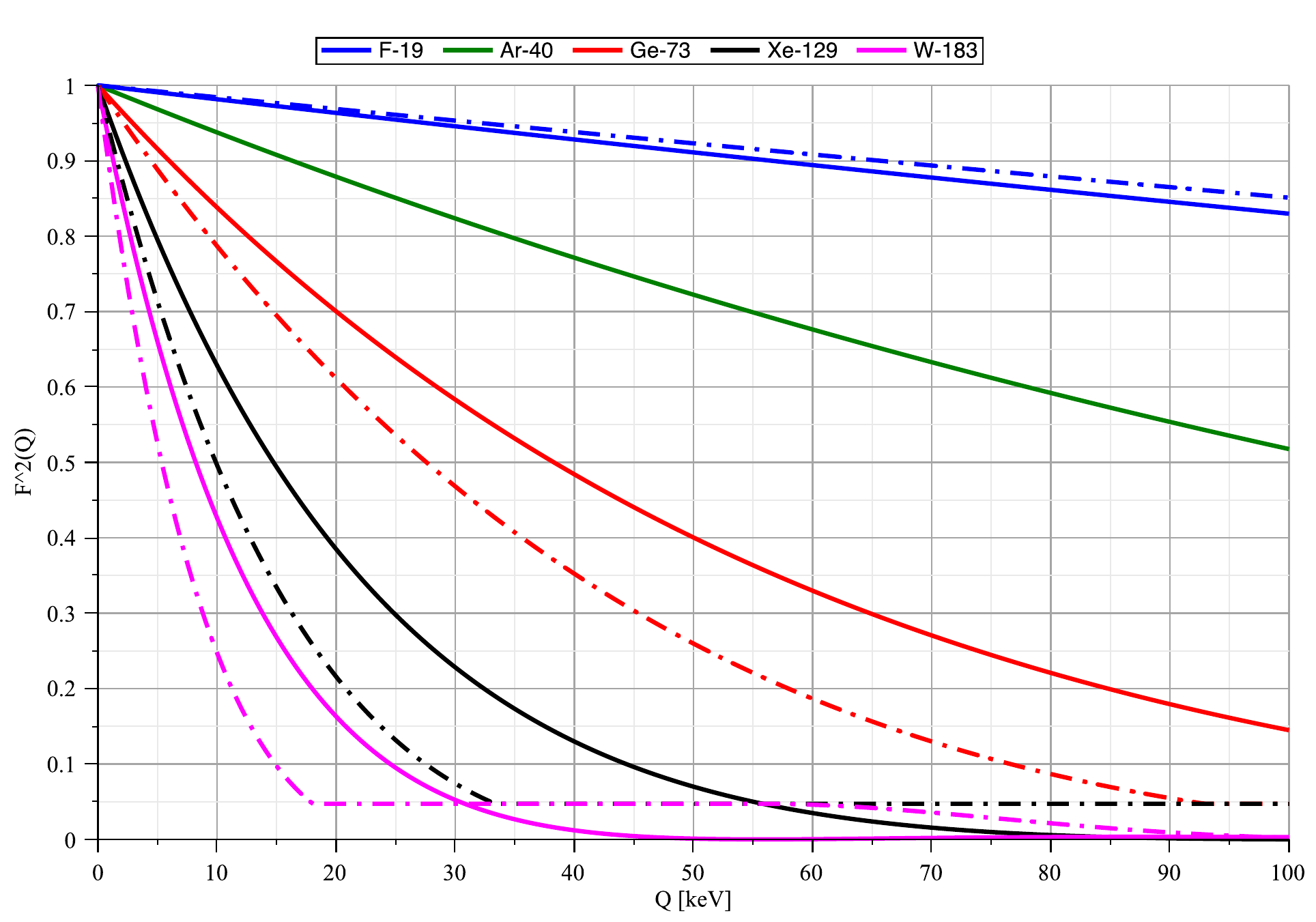}
 \caption{Event number}
 \end{subfigure}
 \hspace{0.1 cm}
 \begin{subfigure} [c] {8.25 cm}
  \includegraphics [width = 8.25 cm] {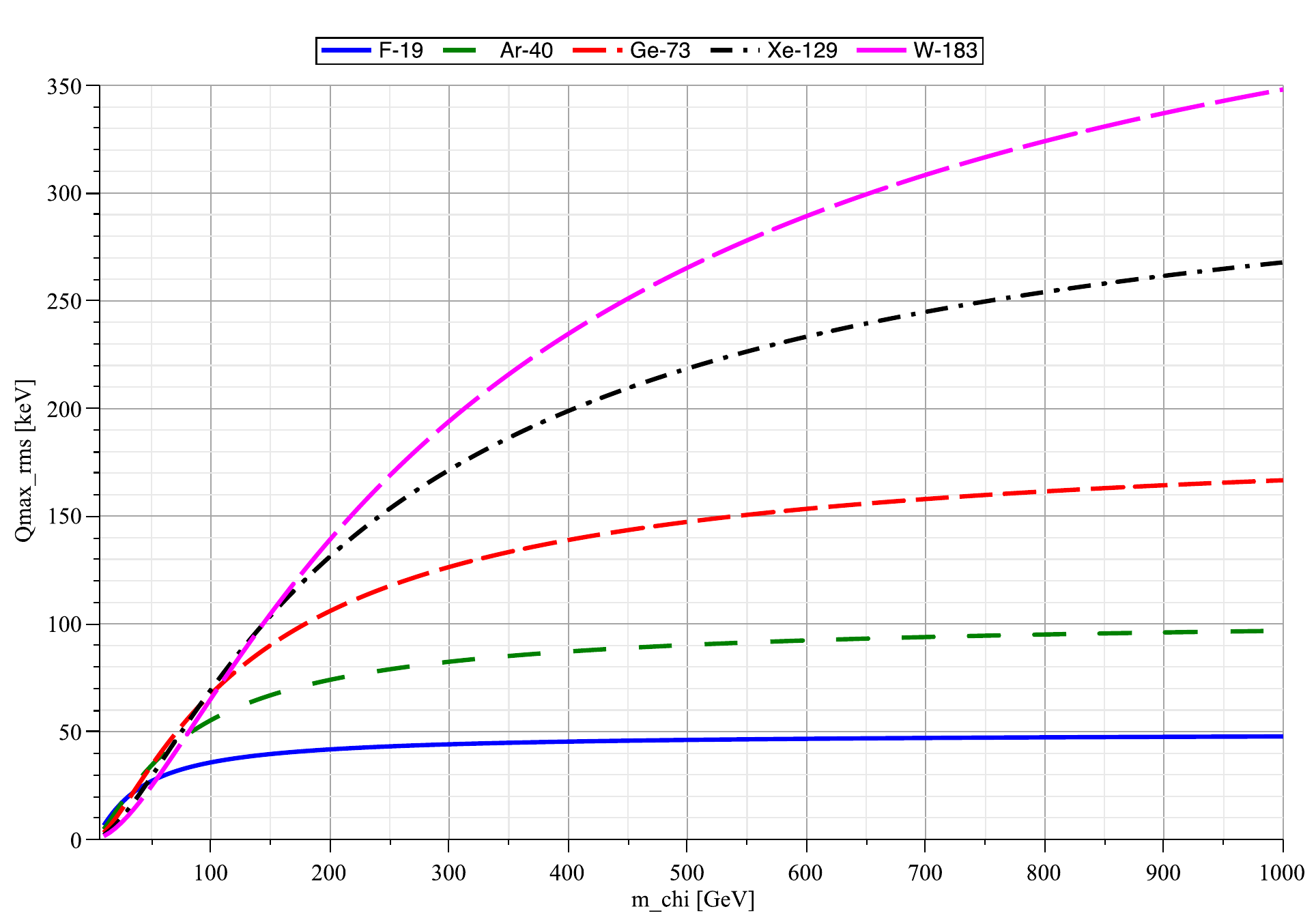}
 \caption{Accumulated recoil energy}
 \end{subfigure}
\end{center}
\caption{
 (a)
 Nuclear form factors of
 the $\rmF$     (blue),
 the $\rmAr$    (green),
 the $\rmGe$    (red),
 the $\rmXe$    (black),
 and the $\rmW$ (magenta) nuclei
 (adopted in our simulation package \cite{DMDDD-3D-WIMP-N})
 as functions of the recoil energy
 between 0 and 100 keV.
 The solid and the dash--dotted curves
 indicate the form factors corresponding to
 the SI and the SD cross sections,
 respectively.
 (b)
 The WIMP--mass dependence of
 the maximal transferable recoil energy
 estimated with the root--mean--square velocity
 of incident halo WIMPs
 in the laboratory reference frame
 $v_{\rm rms, Lab} \simeq 355$ km/s
 \cite{DMDDD-3D-WIMP-N}
 up to $\mchi = 1$ TeV.
 The same five nuclei
 are shown:
 $\rmF$     (solid        blue),
 $\rmAr$    (rare--dashed green),
 $\rmGe$    (dashed       red),
 $\rmXe$    (dash--dotted black),
 and $\rmW$ (long--dashed magenta).
 (Figures from Ref.~\cite{DMDDD-fv_eff}).
}
\label{fig:FQ}
\end{figure}

 As discussed in detail
 in Ref.~\cite{DMDDD-fv_eff},
 for WIMPs moving with low (high) incident velocities
 and thus carrying small (large) kinetic energies,
 the maximal transferable recoil energies to target nuclei
 are small (could be pretty large).
 Then
 the cross section (nuclear form factor) suppression
 and in turn
 the reduction of the scattering probability
 are weak or even negligible (could be pretty strong),
 especially
 when target nuclei and/or WIMPs are light (heavy)
 (See Figs.~\ref{fig:FQ}
  for a direct comparison).

 Consequently,
 ...

\subsubsection*{Forward--backward asymmetry of the 3-D WIMP effective velocity distribution}
\label{sec:FBA}

 In a general ``3-dimensional'' point of view,
 WIMPs moving (approximately)
 in the same (opposite) direction as (of)
 the Galactic movement of our Solar system
 (more precisely,
  of the Earth)
 would have smaller (higher) relative incident velocity
 to the Earth/laboratory.
 Now consider
 the superposition of
 the flux proportionality
 to the WIMP incident velocity
 and
 the nuclear form factor suppression.
 Once WIMPs are as light as only a few tens GeV,
 since 
 the factor of
 the flux proportionality
 dominates,
 the forwardly--moving WIMPs
 would have smaller probabilities
 to scatter off
 both of light and heavy target nuclei;
 in contrast,
 once WIMPs are as heavy as a few hundreds GeV,
 the factor of
 the nuclear form factor suppression
 becomes dominated
 and
 the forwardly--moving WIMPs
 would have larger probabilities
 to scatter off
 heavy target nuclei
 \cite{DMDDD-fv_eff}.
 This can be observed clearly
 from the angular distribution patterns of
 the 3-D Galactic velocity of WIMPs
 {\em scattering} off different target nuclei 
 and thus
 was named as
 the ``forward--backward asymmetry'' of
 the 3-D WIMP effective velocity distribution
 \cite{DMDDD-fv_eff}.

 Moreover,
 our simulations presented in Ref.~\cite{DMDDD-fv_eff}
 showed that,
 once both of WIMPs and target nuclei are light,
 the average incident velocity of scattering WIMPs
 are larger than
 the average velocity of entire halo WIMPs.
 With the increasing WIMP or target mass,
 the average incident velocity of scattering WIMPs
 would be somehow reduced
 to be smaller than
 the average of entire halo WIMPs.

\subsubsection*{Two types of the annual modulation of the 1-D WIMP effective velocity distribution}
\label{sec:annual_fv_eff}

 As mentioned above,
 it is well known that,
 due to the Earth's orbital motion around the Sun,
 the Earth's velocity
 relative to the Dark Matter halo
 is slightly larger (smaller)
 in summer (winter).
 Accompanied with
 the forward--backward asymmetry of
 the WIMP effective velocity distribution,
 this implies that
 WIMPs moving (approximately)
 in the opposite direction of
 the Galactic movement of the Earth
 would have a--bit--much larger (smaller) relative (incident) velocity
 to the Earth/laboratory
 in summer (winter),
 while
 the relative (incident) velocity of
 WIMPs moving (approximately)
 in the same direction as
 the Earth's Galactic movement
 would averagely be almost unchanged.

 Hence,
 in Ref.~\cite{DMDDD-fv_eff}
 we found interestingly that,
 for cases that
 WIMPs or our target nuclei are as light as only a few tens GeV,
 the average Galactic velocity of
 the scattering WIMPs
 would be minimal (maximal)
 in summer (winter),
 whereas
 once both of WIMPs and our target nuclei are as heavy as a few hundreds GeV,
 the average Galactic velocity of
 the scattering WIMPs
 would inversely become maximal (minimal)
 in summer (winter).

%
%

%
% 2/4
 %
%
\section{Simulation results}
\label{sec:results}

 In this section,
 we follow basically our earlier works to numerically simulate
 full 3-D elastic WIMP--nucleus scattering process event--by--event
 \cite{DMDDD-3D-WIMP-N, DMDDD-NR}:
 we generate first a {\em 3-dimensional velocity} of
 an incident WIMP
 in the {\em Galactic} coordinate system
 according to the theoretical isotropic Maxwellian velocity distribution,
 transform it to the laboratory coordinate system,
 and,
 in the laboratory
 (more precisely,
  the incoming--WIMP) coordinate system,
 we generate an {\em equivalent recoil angle} of
 a scattered target nucleus
 and validate this candidate scattering event
 according to the criterion
 \cite{DMDDD-3D-WIMP-N}:
\begin{equation}
     f_{\rm N_R}(\vchiLab, \thetaNRchi)
  =  \frac{\vchiLab}{v_{\chi, {\rm cutoff}}}
     \bigg[\sigmaSI \FSIQ + \sigmaSD \FSDQ\bigg]
     \sin(2 \thetaNRchi)
\,.
\label{eqn:f_NR_thetaNRchi}
\end{equation}
 Here $\vchiLab$ and $\thetaNRchi$ are
 the transformed WIMP incident velocity
 in the laboratory coordinate system
 and the generated equivalent recoil angle of
 the scattered target nucleus%
\footnote{
 The elevation of the nuclear recoil direction
 in the incoming--WIMP coordinate system
 and the complementary angle of the recoil angle.
},
 $v_{\chi, {\rm cutoff}}$
 is a cut--off velocity of incident halo WIMPs
 in the Equatorial/laboratory coordinate systems,
 which is set as 800 km/s
 in our simulations,
 $\sigma_0^{\rm (SI, SD)}$ are
 the spin--independent/dependent (SI/SD) total cross sections
 ignoring the nuclear form factor suppressions,
 and
 $F_{\rm (SI, SD)}(Q)$ indicate
 the elastic nuclear form factors
 corresponding to the SI/SD WIMP interactions,
 respectively.
 For the accepted scattering events,
 we record the numbers and the accumulated recoil energies
 in different advanced seasons
 \cite{DMDDD-3D-WIMP-N}.

 Four spin--sensitive nuclei:
 $\rmF$,
 $\rmGe$,
 $\rmXe$,
 and $\rmW$
 have been considered as our targets,
 so that
 our simulation results could basically cover
 the mass range of almost all nuclei
 used in direct DM detection experiments.
 And,
 as in our earlier works
 presented in Refs.~%
 \cite{DMDDD-NR, DMDDD-fv_eff},
 the SI (scalar) WIMP--nucleon cross section
 has been fixed as $\sigmapSI = 10^{-9}$ pb,
 while
 the effective SD (axial--vector) WIMP--proton/neutron couplings
 have been tuned as $\armp = 0.01$
 and $\armn = 0.7 \armp = 0.007$,
 respectively.
 So that
 the contributions of
 the SI and the SD WIMP--nucleus cross sections
 (including the corresponding nuclear form factors)
 in Eq.~(\ref{eqn:f_NR_thetaNRchi})
 are approximately comparable:
\begin{equation}
        \sigmaSI \FSIQ
  \sim  \sigmaSD \FSDQ
%\~.
\label{eqn:sigmaSI/SD_sim}
\end{equation}
 for $\rmGe$ and $\rmXe$ nuclei%
\footnote{
 However,
 for $\rmF$ and $\rmW$ nuclei,
 since their masses are either too light or too heavy,
 the SD or the SI WIMP--nucleus cross section dominates.
}.

 Moreover,
 in this paper
 we assume simply that
 the threshold energies
 for all considered target nuclei
 are negligible,
 whereas
 the maximal experimental cut--off energy
 has been set as $\Qmax = 100$ keV.
\footnote{
 Note that,
 for cases with non--negligible threshold energies
 and/or narrower energy windows
 (e.g., 20 or \mbox{30 keV}),
 the WIMP--mass and target dependent annual modulation
 would be more complicated.
}
 5,000 experiments
 with 5,000 {\em accepted} events on average
 (Poisson--distributed)
 in one entire year
 in one experiment
 have been simulated.

\subsection{Normal annual modulation}
\label{sec:normal}

 We consider two cases
 performing the normal annual modulation
 at first.

\subsubsection*{\boldmath
                100-GeV WIMPs off $\rmF$ target nuclei}
\label{sec:F19-0100}
 \def \Target   {F19}
 \def \WIMPmass {0100}
 \InsertPlotNQ
  {The event number (a)
   and the accumulated recoil energy (b) of
   $\rmF$ target nuclei
   scattered by 100-GeV WIMPs.
   5,000 accepted
   WIMP scattering events on average
   (Poisson--distributed)
   in one entire year
   in one experiment
   have been simulated
   and binned into 6 bins.
   The dashed blue vertical bars indicate
   the 1$\sigma$ statistical uncertainties,
   while
   the dash--dotted red horizontal lines indicate
   the yearly average value.%
   }

 In Figs.~\ref{fig:NQ-F19-0100-5000-04949-00},
 we show
 the event number (a)
 and the accumulated recoil energy (b) of
 $\rmF$ target nuclei
 scattered by 100-GeV WIMPs.
 The dashed blue vertical bars indicate
 the 1$\sigma$ statistical uncertainties,
 while
 the dash--dotted red horizontal lines indicate
 the yearly average value.

 As expected,
 the event numbers in different seasons
 show a sinusoidal time dependence.
 Only unfortunately,
 due to the relatively small event numbers
 (833 events/bin on average),
%
% 833.33 $\pm$ 28.87
% 833.15 $\pm$ 28.86
%
 the (dashed blue) statistical error bars
 are still too large to be clearly distinguished from each other.
 Meanwhile,
 the seasonal variation of
 the accumulated recoil energies
 in Frame (b)
 shows a relatively better sinusoidal modulation
 with an $\sim$ 7.1\% ($\sim 1.56\sigma$) variation amplitude.
%
% 856.0538 - 833.1547 = 22.8991,   22.8991 / 833.1547 = 0.0275,   22.8991 / sqrt(856.0538) = 0.783
% 833.1547 - 809.8322 = 23.3225,   23.3225 / 833.1547 = 0.0280,   23.3225 / sqrt(809.8322) = 0.820
%
% 17948.55 - 16745.56 = 1202.99,   1202.99 / 16745.56 = 0.0718,   1202.99 / 812.3053 = 1.48
% 16745.56 - 15551.59 = 1193.97,   1193.97 / 16745.56 = 0.0713,   1193.97 / 732.9121 = 1.63
%

%
%
\subsubsection*{\boldmath
                20-GeV WIMPs off $\rmXe$ target nuclei}
\label{sec:Xe129-0020}
 \def \Target   {Xe129}
 \def \WIMPmass {0020}
 \InsertPlotNQ
  [b!]
  {As Figs.~\ref{fig:NQ-F19-0100-5000-04949-00},
   except that
   a heavy nucleus $\rmXe$
   has been considered as our target
   and
   the WIMP mass has been lowered to only 20 GeV.%
   }

 In Figs.~\ref{fig:NQ-Xe129-0020-5000-04949-00},
 we consider
 a heavy nucleus $\rmXe$
 as our target
 and
 the WIMP mass has been lowered to only 20 GeV.
 Again,
 while
 the seasonal variation of
 the event number
 shows a sinusoidal time dependence
 with large statistical uncertainties,
 that of
 the accumulated recoil energies
 shows a somehow better sinusoidal modulation
 with an $\sim$ 6.3\% ($\sim 1.35\sigma$) variation amplitude.
%
% 850.7358 - 833.6969 = 17.0389,   17.0389 / 833.6969 = 0.0204,   17.0389 / sqrt(850.7358) = 0.584
% 833.6969 - 816.6264 = 17.0705,   17.0705 / 833.6969 = 0.0205,   17.0705 / sqrt(816.6264) = 0.597
%
% 3021.262 - 2839.802 = 181.460,   181.460 / 2839.802 = 0.0639,   181.460 / 139.8580 = 1.30
% 2839.802 - 2660.752 = 179.050,   179.050 / 2839.802 = 0.0631,   178.710 / 126.7110 = 1.41
%

%
%

%
% 3/4
 %
%
\subsection{Reverse annual modulation}
\label{sec:reverse}

 Now
 we turn to consider two cases of
 heavy WIMPs
 scattering off heavy target nuclei.

\subsubsection*{\boldmath
                500-GeV WIMPs off $\rmXe$ target nuclei}
\label{sec:Xe129-0500}
 \def \Target   {Xe129}
 \def \WIMPmass {0500}
 \InsertPlotNQ
  {As Figs.~\ref{fig:NQ-Xe129-0020-5000-04949-00}:
   the heavy nucleus $\rmXe$
   has been considered,
   but
   the WIMP mass has been raised to 500 GeV.%
   }

 In Figs.~\ref{fig:NQ-Xe129-0500-5000-04949-00},
 we still use
 the heavy nucleus $\rmXe$
 as our target,
 but
 raise the WIMP mass to \mbox{500 GeV}.
 As expected,
 both of the event number and the accumulated recoil energy
 show roughly {\em reverse} sinusoidal variations.
 However,
 due to larger recoil energies
 (transferred from heavy incident WIMPs)
 and in turn larger statistical uncertainties,
 the $\sim 0.83\sigma$ ($\sim$ 2.9\%) amplitude of
 the seasonal variation of
 the event number
 would now be a (relatively) better indicator.
%
% 833.6063 - 810.0768 = 23.5295,   23.5295 / 833.6063 = 0.0282,   23.5295 / sqrt(810.0768) = 0.827
% 857.8062 - 833.6063 = 24.1999,   24.1999 / 833.6063 = 0.0290,   24.1999 / sqrt(857.8062) = 0.826
%

%
%
\subsubsection*{\boldmath
                200-GeV WIMPs off $\rmW$ target nuclei}
\label{sec:W183-0200}
 \def \Target   {W183}
 \def \WIMPmass {0200}
 \InsertPlotNQ
  [b!]
  {As Figs.~\ref{fig:NQ-Xe129-0500-5000-04949-00},
   except that
   an even heavier nucleus $\rmW$
   has been considered as our target
   and
   the WIMP mass has been lowered to 200 GeV.%
   }

 As a second example,
 in Figs.~\ref{fig:NQ-W183-0200-5000-04949-00},
 we consider
 an even heavier nucleus $\rmW$
 as our target,
 but
 lower the WIMP mass to 200 GeV.
 Again,
 we can see two reverse sinusoidal variation curves
 and
 the modulation amplitude of the event number
 is $\sim 0.85\sigma$ ($\sim$ 2.9\%).
%
% 832.9560 - 808.9528 = 24.0032,   24.0032 / 832.9560 = 0.0288,   24.0032 / sqrt(808.9528) = 0.844
% 858.0950 - 832.9560 = 25.1390,   25.1390 / 832.9560 = 0.0302,   25.1390 / sqrt(858.0950) = 0.858
%

%
%

%
% 4/4
 %
%
\subsection{Intermediate WIMP/target mass}
\label{sec:intermediate}

 Combining results shown previously,
 it would be reasonable to expect that
 there should be
 a (target--dependent) turning point on the WIMP mass
 (more precisely,
  a turning boundary in the parameter space of
  the WIMP mass and different interaction couplings),
 around which
 the annual modulation of
 the WIMP scattering event rate
 would disappear.
 To demonstrate this prediction,
 we consider
 at the end
 two cases with an intermediate WIMP or target mass
 \cite{DMDDD-fv_eff}.

\subsubsection*{\boldmath
                500-GeV WIMPs off $\rmGe$ target nuclei}
\label{sec:Ge73-0500}
 \def \Target   {Ge73}
 \def \WIMPmass {0500}
 \InsertPlotNQ
  {As Figs.~\ref{fig:NQ-Xe129-0500-5000-04949-00}:
   the WIMP mass has been considered as 500 GeV,
   except that
   a middle--mass nucleus $\rmGe$
   has been considered as our target.
   }

 At first,
 in Figs.~\ref{fig:NQ-Ge73-0500-5000-04949-00},
 the WIMP mass has been assumed as heavy as 500 GeV,
 but
 a middle--mass nucleus $\rmGe$
 has been considered as our target.
 Kind of as expected,
 while
 the event numbers in different seasons
 might somehow show a reverse sinusoidal time dependence
 with however
 a very small variation amplitude (only $\sim$ 1.5\%),
 the seasonal variation of
 the accumulated recoil energy
 would rather be uniform.
%
% 832.9109 - 820.1932 = 12.7177,   12.7177 / 832.9109 = 0.0153,   12.7177 / sqrt(820.1932) = 0.444
% 845.3414 - 832.9109 = 12.4305,   12.4305 / 832.9109 = 0.0149,   12.4305 / sqrt(845.3414) = 0.428
%

%
%
\subsubsection*{\boldmath
                100-GeV WIMPs off $\rmXe$ target nuclei}
\label{sec:Xe129-0100}

 As a second example,
 in Figs.~\ref{fig:NQ-Xe129-0100-5000-04949-00},
 we use again
 the heavy nucleus $\rmXe$
 as our target,
 but
 assume an intermediate WIMP mass of 100 GeV.
 Similarly,
 while
 the seasonal variation of
 the event number
 show a reverse sinusoidal time dependence
 with an even smaller variation amplitude (only $\sim$ 1.4\%),
 that of
 the accumulated recoil energy
 show a sinusoidal time dependence
 with an $\sim$ 1.1\% variation amplitude.
%
% 833.1754 - 821.2892 = 11.8862,   11.8862 / 833.1754 = 0.0143,   11.8862 / sqrt(821.2892) = 0.415
% 844.6098 - 833.1754 = 11.4344,   11.4344 / 833.1754 = 0.0137,   11.4344 / sqrt(844.6098) = 0.393
%
% 11420.79 - 11306.61 = 114.18,   114.18 / 11306.61 = 0.0101,   114.18 / 566.3084 = 0.202
% 11306.61 - 11172.13 = 134.48,   134.48 / 11306.61 = 0.0119,   134.48 / 546.5217 = 0.246
%

%
 \def \Target   {Xe129}
 \def \WIMPmass {0100}
 \InsertPlotNQ
  {As Figs.~\ref{fig:NQ-Xe129-0020-5000-04949-00}
   and \ref{fig:NQ-Xe129-0500-5000-04949-00}:
   the heavy nucleus $\rmXe$
   has been considered,
   but
   the WIMP mass has been chosen as 100 GeV.%
   }
\section{Summary}

 In this paper,
 following our earlier work on
 the 3-dimensional effective velocity distribution of
 Galactic WIMPs
 scattering off target nuclei,
 we demonstrated
 the normal and the reverse annual modulations of
 elastic WIMP--nucleus scattering signals,
 which could be observed
 in direct Dark Matter detection experiments.

 Our simulations show that,
 once the WIMP mass is as light as only a few tens GeV,
 the event number and the accumulated recoil energy of
 WIMP--induced scattering events off
 both of light and heavy target nuclei
 would indeed be maximal (minimal)
 in summer (winter).
 With $\cal O$(5,000) recorded events
 in one or several consecutive years,
 the amplitude of the normal annual modulation of
 the accumulated recoil energy
 would be $\sim 1.35\sigma$ (Xe)
 to $\sim$ 1.56$\sigma$ (F)
 (depending on
  the target and the WIMP masses
  as well as
  the analyzed energy window).

 However,
 once the WIMP mass is as heavy as a few hundreds GeV,
 the event number and the accumulated recoil energy of
 WIMP scattering events off heavy nuclei
 would inversely be minimal
 in summer.
 With $\cal O$(5,000) recorded events,
 the amplitude of the reverse annual modulation of
 the event number
 could be smaller than $\sim 1\sigma$.
 For an intermediate WIMP mass,
 our simulations demonstrated that
 the event number and the accumulated recoil energy of
 scattering events off
 some middle--mass nuclei
 would show an approximately uniform
 or very tiny seasonal variation.

 In summary,
 by applying our simulation package
 for 3-D elastic WIMP--nucleus scattering,
 we studied
 the annual modulation of WIMP scattering signals
 under a different approach.
 Hopefully,
 what we observed in this work
 could be helpful to our colleagues
 in analyzing experimental data.

\subsubsection*{Acknowledgments}

 This work
 was strongly encouraged by
 the ``{\it Researchers working on
 e.g.~exploring the Universe or landing on the Moon
 should not stay here but go abroad.}'' speech.

%
%
% References
%
 %
%

%
%

%

%
%
%
\end{document}